# Time-reversal symmetry breaking versus superstructure

Sergey V. Borisenko[1], Alexander A. Kordyuk[1,2], Andreas Koitzsch[1], Martin Knupfer[1], Jörg Fink[1].

[1]*Institute for Solid State Research, IFW-Dresden, P.O. Box 270016, D-01171, Dresden, Germany*

[2]*Institute of Metal Physics, 03142 Kyiv, Ukraine*

**One of the mysteries of modern condenced-matter physics is the nature of the pseudogap state of the superconducting cuprates. Kaminski et al.[1] claimed to have observed signatures of time-reversal symmetry breaking in the pseudogap regime in underdoped $Bi_2Sr_2CaCu_2O_{8+\delta}$ (Bi2212). Here we argue that the observed dichroism is due to the 5x1 superstructure replica of the electronic bands and therefore cannot be considered as evidence for the spontaneous time-reversal symmetry breaking in cuprates.**

The main conclusions of Kaminski et al. are based on the temperature dependent circular dichroism observed in a "mirror" plane of the underdoped Bi2212. However, pristine Bi2212 samples possess an incommensurate modulation of the Bi-O layers resulting in an approximately 5x1 superstructure along the crystallographic **b** direction. In Figure 1a, b (left panels) we show angular distributions of elastically scattered electrons and photoelectrons as seen in Low Energy Electron Diffraction (LEED) and Angle-Resolved Photoemission (ARPES) experiments, respectively. Both experiments clearly indicate the absence of reflection symmetry in the planes corresponding to the Cu-O bonds (white dotted lines in Figure 1a), which are thus not mirror planes. In order to remove the 5x1 superstructure one may dope the pristine Bi2212 with lead. As a result, the planes in question become mirror planes, as demonstrated in the right panels of Figure 1a, b.

We have performed ARPES experiments similar to those reported in Ref.1 using circularly polarized light on both systems. Already at room temperature (Figure 1c) the influence of the superstructure is obvious: for pristine Bi2212 the dichroic signal is not symmetric with respect to the $\Gamma$ - $(\pi, 0)$ plane and is non-zero. This result, although in direct contradiction with the data of Kaminski et al., is expected and easy to understand. Superstructure results in diffraction replicas of the electronic structure seen in the momentum distribution map (Figure 1b) and schematically shown in energy-momentum coordinates (Figure 1d) as green and blue dashed curves. It is well known that due to the pronounced inequivalence of the matrix elements in the first and second Brillouin zones the spectral weight of these replicas is always different near the $(\pi, 0)$-point. In this particular case, the "blue" one is apparently stronger than the "green". Recording the dichroism as a function of **k** along the white arrow (Figure 1b) one effectively measures the superposition of the three signals originating from the main band and two non-equivalent diffraction replicas. A systematic investigation of the 5x1 superstructure-free Pb-Bi2212 samples shows that the dichroism in the mirror plane remains zero within the experimental error bars independent of temperature and doping[2].

In spite of the severe quantitative discrepancy (partially caused by the use of different photon energies) qualitative agreement between our data and the data of Kaminski et al. taken on pure Bi2212 can be achieved assuming that the zero of their momentum scale does not correspond to the $(\pi, 0)$-point. In contrast to our study[2], where the determination of the momentum scale plays a central role, nothing is said about this demanding procedure in Ref. 1. The uncertainty in determination of the momentum scale can be estimated from other data shown in their paper. EDCs presented in Fig.3 (Ref.1) for the overdoped sample are claimed to be $k_F$-EDCs. It is known however, that at finite temperature at $k_F$ the spectral function has a peak at the chemical potential and multiplication by the Fermi function would result in the leading edge midpoint located at negative binding energies, which is clearly not the case there. Our quantitative estimates[3] show that presented EDC's are taken more than 0.02 Å away from the $k_F$, thus probably representing the typical accuracy of the momentum scale determination in Ref.1.

Provided the zero momentum in Fig. 3g of Ref.1 does not correspond to the (π, 0)-point, the temperature dependence of the dichroism is not surprising at all. Away from the mirror plane already the dichroism corresponding to the main band is temperature dependent[2]. This is also seen in Fig. 3c of Ref.1 (note, that the lines shown in Figs. 3 c and g are not always linear fits to 11 data points, as is evident for the 150 K data). Absence of full-range curves[2] in Figs.3 c, g does not allow determining where exactly in momentum space the presented data are taken and whether this is always the same place. In favor of this is the considerable (~14%) variation of the slope of the 250 K "line" in two similarly underdoped samples (as shown in Fig. 3g and Fig. 4a, b) which cannot be explained by the small difference in doping levels since the comparison with a much stronger doped sample (see Fig. 3 c) gives a comparable change of the slope.

The data presented in Fig. 4 a, b (Ref. 1) also fit the "superstructure scenario" – as follows from Figure 1, when going from (π, 0) to (0, π) point, the stronger diffraction replica is now on the other side of the (0, π)-point and therefore the temperature induced changes may naturally have the opposite sign.

Finally, the data on the overdoped sample (Fig. 3 c in Ref. 1) can be explained by the substantially weaker influence of the diffraction replica in the immediate vicinity of the (π, 0) – point because of the larger size of the Fermi surface (Figure 1d). In addition, the superlattice signal seems to be more sensitive to the temperature in underdoped samples[4] and vary from sample to sample: in Ref. 1 it is reported to be around 3% while it is seen in Ref. 5 to be ~10%.

These arguments supported by experimental data demonstrate that the observed dichroism in the mirror plane is due to the 5x1 superstructure in pristine Bi2212.

1. Kaminski, A. et al. *Nature* **416,** 610–613 (2002).

2. Borisenko, S. V. et al. cond-mat/0305179.

3. See, e.g., Kordyuk, A. A. et al. Phys. Rev. B 67 064504 (2003).

4. Armitage, N. P. and J. Hu, Phil. Mag. Lett. (in press), cond-mat/0303186.

5. Kaminski, A et al. cond-mat/0306140.

Correspondence and requests for materials should be addressed to SVB. (e-mail: S.Borisenko@ifw-dresden.de).

Dichroism due to the superstructure. a, LEED and b, ARPES angular distributions of the electrons in pristine (left) and Pb-doped (right) Bi2212. White dotted lines represent crystallographic planes. Green and blue dashed lines – diffraction replicas. White dashed line – first Brillouin zone. c, Room temperature dichroism near ($\pi$, 0) in Bi2212 and Pb-Bi2212. d, diffraction replicas in (E, k)-coordinates.

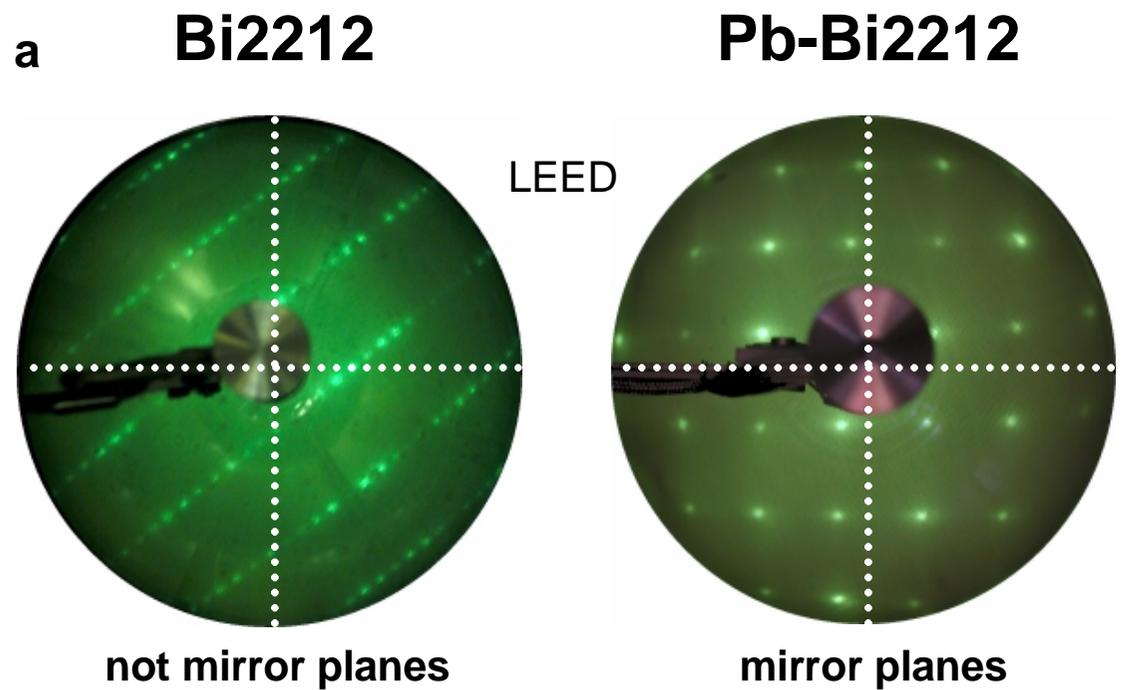
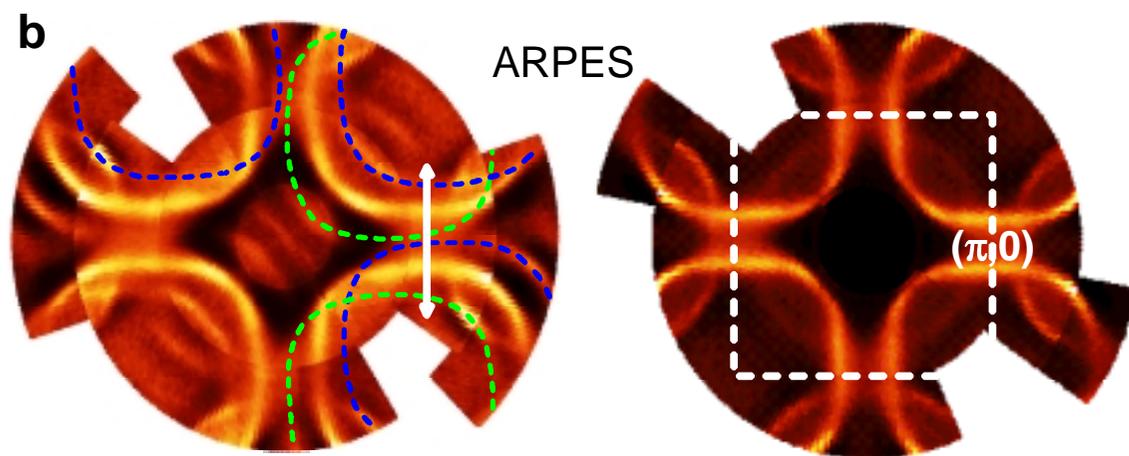
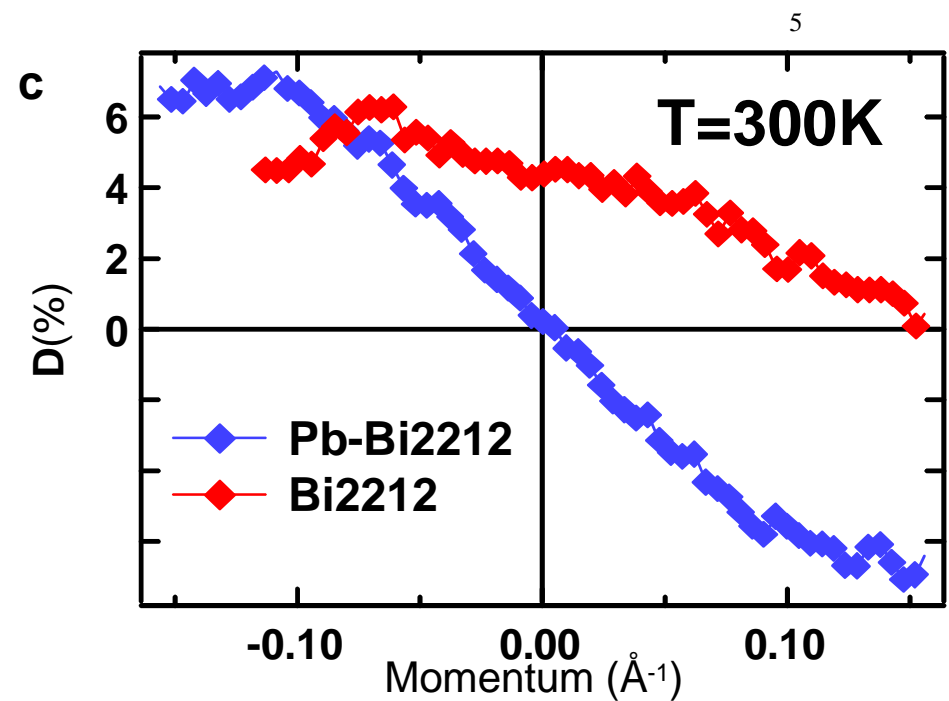
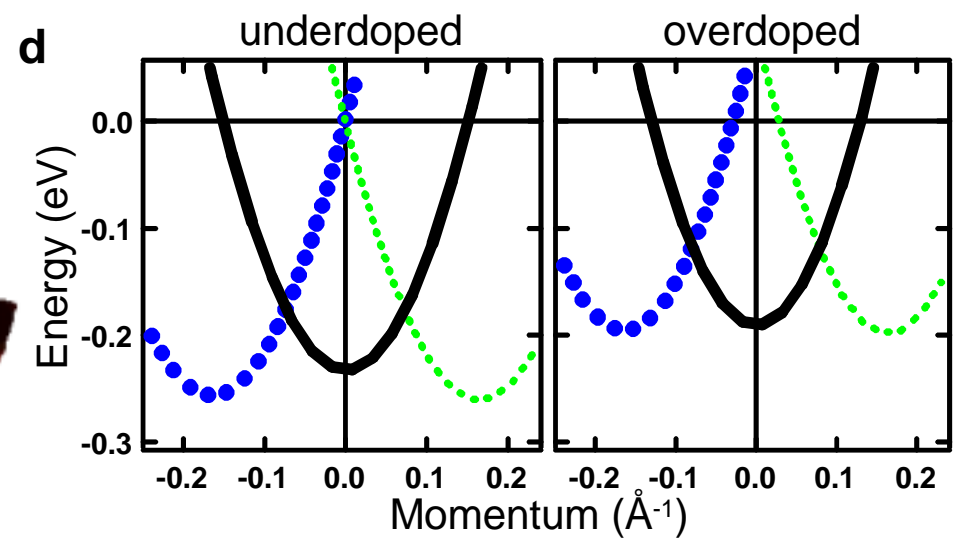